\documentclass[12pt]{iopart}
\usepackage{amssymb,graphicx}

\makeatletter

\@addtoreset{equation}{section} \makeatother

\newcommand{\be}{\begin{equation}}
\newcommand{\ee}{\end{equation}}
\newcommand{\bea}{\begin{eqnarray}}
\newcommand{\eea}{\end{eqnarray}}
\newcommand{\gs}{\delta_{\scriptscriptstyle \rm GS}}
\newcommand{\vp}{\varphi}
\renewcommand{\Re}{\,\mathrm{Re}\,}
\renewcommand{\Im}{\,\mathrm{Im}\,}

\begin{document}

\title{\bf Moduli Corrections to $D$-term Inflation}

\author{Ph~Brax$^1$, C~van~de~Bruck$^2$, A~C~Davis$^3$,
S~C~Davis$^4$, R~Jeannerot$^4$ and M~Postma$^5$}

\address{ ${}^1$ Service de Physique Th\'eorique, CEA/DSM/SPhT,
Unit\'e de recherche associ\'ee au CNRS, CEA--Saclay
F--91191 Gif/Yvette cedex, France}
\address{${}^2$ Department of Applied Mathematics, University of Sheffield,
Hounsfield Road, Sheffield S3 7RH, UK}
\address{${}^3$ DAMTP, Centre for Mathematical Sciences,
University of Cambridge, Wilberforce Road, Cambridge,CB3 0WA, UK}
\address{${}^4$ Instituut-Lorentz for Theoretical Physics, Postbus 9506,
 NL--2300 RA Leiden, The Netherlands}
\address{${}^5$ Nikhef, Kruislaan 409, NL--1098 SJ Amsterdam, The
Netherlands}

\eads{\mailto{brax@spht.saclay.cea.fr}, \mailto{C.vandeBruck@sheffield.ac.uk},
\mailto{A.C.Davis@damtp.cam.ac.uk}, \mailto{sdavis@lorentz.leidenuniv.nl},
\mailto{jeannero@lorentz.leidenuniv.nl}, \mailto{mpostma@nikhef.nl}}

\begin{abstract}
We present a $D$-term hybrid inflation model, embedded in
supergravity with moduli stabilisation. Its novel features
allow us to overcome the serious challenges of combining $D$-term
inflation and moduli fields within the same string-motivated
theory. One salient point of the model is the
positive definite uplifting $D$-term arising from
the moduli stabilisation sector. By coupling this $D$-term to the
inflationary sector, we generate an effective Fayet-Iliopoulos term. 
Moduli corrections to the inflationary dynamics are also obtained. 
Successful inflation is achieved for a limited
range of parameter values with spectral index compatible with the WMAP3 data.
Cosmic $D$-term strings are also formed at the end of inflation; these
are no longer Bogomol'nyi-Prasad-Sommerfeld (BPS) objects. The
properties of the strings are studied.

\end{abstract}

\maketitle

\section{Introduction}

Inflation is the favoured solution of the age and flatness problems of
the standard cosmology, and also provides a mechanism for generating
density perturbations and the cosmic microwave background (CMB)
anisotropies. Obtaining a sufficiently flat potential for successful
inflation is difficult, although supersymmetric (SUSY) hybrid
inflation seems to offer an answer~\cite{Cop,Dvasha,Dterm}. Hybrid inflation 
uses three fields, a gauge singlet which is the slow-rolling
field, and two oppositely charged fields which acquire vacuum
expectation values at the end of inflation: hybrid inflation provides
a mechanism to naturally end inflation \cite{hybrid}.  Among the
different realisations of hybrid inflation, $D$-term
inflation~\cite{Dterm} is singled out because it is free from the
$\eta$-problem~\cite{eta}.

The discovery of branes in string theory has opened up new ways of
building realistic inflationary models \cite{review1,review2}.  Hybrid
inflation and in particular supersymmetric hybrid inflation has been
thoroughly investigated within this context. It has a natural
realisation within the D3/D7 system of string theory where the
inflaton becomes the interbrane distance and the waterfall fields
correspond to interbrane open string degrees of freedom~\cite{D3D7}.
$D$-term inflation can be interpreted as an
effective field theory of brane inflation. This
interpretation is strengthened by the fact that BPS cosmic strings
form at the end~\cite{Dstrings}. We find that the
strings are no longer BPS when moduli fields are included, at least
for the stabilisation mechanism considered in this paper.

In string theory, many moduli appear after compactification. These
moduli threaten to destabilise vacua by being either flat
directions or having runaway potentials. Over the last couple of
years great progress has been made towards stabilising the moduli,
probably best exemplified by the KKLT proposal.   In such a
scenario, all moduli are stabilised using a combination of
fluxes and non-perturbative effects~\cite{KKLT}. Obtaining a
Minkowski vacuum requires a lifting procedure for the
supersymmetric AdS vacuum. This can be realised in a
non-supersymmetric fashion with the addition of anti-branes. A
proper embedding within supergravity, where the  lifting is
done by a $D$-term, has been obtained more
recently~\cite{BKQ,nilles,ana}.

In a supergravity theory all fields couple to each other, at
least with gravitational strength, and hence the moduli and inflation
sectors will interact. As has been observed
in~\cite{Lyth}, fields with values near the Planck scale can
destroy the flatness of the inflationary potential.
The moduli can therefore prevent slow-roll inflation. It is also
possible for the inflation vacuum energy to destabilise the moduli.
Studies of supergravity corrections to hybrid inflation have been carried
out in \cite{old1,infl}. However the above two problems were not
studied there, as no specific form for the
moduli sector superpotential was introduced. They were
considered in \cite{Fterm}, where yet another problem was found.
There it was shown that the variation of the moduli during
inflation can prevent slow-roll in
$F$-term inflation, even if the moduli variation is tiny.
Moduli fields therefore pose a serious problem to any attempt to
embed inflationary cosmology within supergravity, even when they
are stable.

Inflation can only work if moduli effects are small.  This is a
far from trivial requirement.   Indeed, keeping the moduli stable
during inflation requires the moduli scale to be larger than the
inflationary scale, whereas the need for small soft masses seems to imply a low
moduli scale. In this paper, we will use a $D$-term hybrid
inflation model~\cite{Dterm}. In contrast to $F$-term inflation,
$D$-term is thought to be free from the $\eta$-problem. We find
that this is no longer true when moduli fields are included. There
is another reason to expect $D$-term inflation to be a better
choice. Many of the major problems of combining $F$-term inflation
with moduli fields arise because both sectors of the theory are
using non-zero $F$-terms, which mix in the full potential. On the
other hand, the $F$-term used in $D$-term inflation is zero during
inflation, and so the effects of its mixing with the moduli sector
are reduced. As we will see, this fact does indeed allow $D$-term
inflation to be successfully combined with moduli stabilisation.

However, the use of $D$-term inflation introduces an extra
complication. The vacuum energy driving inflation is provided by a
Fayet-Iliopoulos (FI-)term. It is an old result that in $N=1$
supergravity an FI-term  can only be present when there exists an 
R-symmetry~\cite{freedman}. This requires that the full
superpotential, including terms from other, non-inflationary
sectors, is charged. Although not necessarily impossible to achieve, this
does complicate model building enormously.  To get around this problem
we will generate an effective FI-term from the VEVs of other fields.
The most economic solution is to use the moduli fields.
By charging them under the inflationary $U(1)$, their VEVs will
provide the FI-term needed, which is constant in the limit that the
moduli are fixed during inflation.

In this paper, the stabilisation of moduli with a Minkowski vacuum is
achieved thanks to an Abelian gauge symmetry differing from the gauge
symmetry of the $D$-term inflation sector.  A key element in our model
is the K\"ahler potential, which is of the no-scale form, with all
matter fields inside the logarithm.  As a result the soft masses are
much smaller than what might naively be inferred from the SUSY
breaking scale. Moreover, a shift symmetry is introduced to keep the
inflaton potential flat~\cite{braxshift}.

The FI-term is not a free parameter, but is fixed by the COBE
normalisation of the density perturbations produced during
inflation. In addition, the requirement of a red tilted spectrum
further constrains the model parameters.  We find that inflation
works if the moduli corrections to the standard SUSY $D$-term
inflation results are small. In a large part of the parameter
space, the moduli corrections  strongly depend on the cutoff
scale, and a RGE analysis and/or 2nd order
calculation is needed to make precise statements. The tendency
is that they lower the spectral index, which can give a better
agreement with the latest WMAP3 result~\cite{WMAP3}. In addition
there is a (fine-tuned) region of parameter space where the moduli effects can
dominate, and the model still gives an acceptable spectral index.

Another notable feature of $D$-term hybrid inflation is the
formation of cosmic strings at the end of inflation~\cite{prd}. In
supersymmetric theories, these $D$-term strings are
BPS~\cite{DDT}. Both strings and inflation contribute to CMB
anisotropies, and current data constrain the string contribution
to be less than a few percent~\cite{pogosian,bevis}. This can only be
satisfied for tiny gauge couplings when the K\"ahler potential is
minimal~\cite{infl,mairi}. Such small gauge couplings are
difficult to reconcile within string theory. Interestingly enough,
with a shift symmetry, the restriction on the gauge couplings can
be evaded; then only tiny superpotential couplings are needed.
Another  way of  avoiding   the cosmic string bound has
been proposed in~\cite{UAD}. If the model is augmented by a second
set of Higgs fields, which are put in a global $SU(2)$ multiplet
together with the original Higgs fields, the strings formed at the
end of inflation are semi-local and will in general
decay~\cite{semilocal}. In addition, SUSY implies the existence of
fermion zero modes on the strings~\cite{DDT}, although these can
disappear when SUGRA is included~\cite{RMstr}. In a $D$-term
inflation model there is an additional inflatino zero mode, which
remains in SUGRA~\cite{RMstr}. The existence of zero modes may
alter the cosmological properties of string loops~\cite{BCDT,CD}.
As with inflation itself, the inclusion of the moduli sector
changes the BPS properties of $D$-term cosmic strings and their
associated zero mode spectrum.

The paper is organised as follows. In section~\ref{sec:review}, we
review standard $D$-term inflation, and describe the uplifting
$D$-term moduli stabilisation mechanism.  In
section~\ref{sec:combo} we discuss the problems encountered when
one tries to combine $D$-term inflation with moduli stabilisation.
We present an explicit model that overcomes all of them. We show
how important the choice of the K\"ahler potential is. The scalar
potential is identically flat at tree level. The one-loop
radiative corrections which give the required slope for slow-roll
inflation are calculated in section~\ref{sec:effepo}. This is
followed by a study of the inflationary dynamics in
section~\ref{sec:inf}. We  consider both  the limit in
which the standard SUSY results are recovered, as well as the
limit in which moduli corrections dominate. Finally, in
section~\ref{sec:str} we discuss the properties of the cosmic
strings formed in our model, and the fermion bound states.  We
give some concluding remarks in section~\ref{sec:conc}.

\section{Background}
\label{sec:review}

\subsection{$D$-term inflation}

We start with a review of standard hybrid inflation in a globally
supersymmetric theory.  $D$-term inflation relies on the existence of 
an Abelian gauge symmetry $U(1)_1$, with a constant
Fayet-Iliopoulos term. The model contains two conjugate superfields
$\phi^+$ and $\phi^-$ charged under this $U(1)_1$, and a gauge singlet
inflaton field $\phi$.  The superpotential is
\be
W_{\rm inf} = \lambda \phi \phi^+ \phi^-.
\label{Winf}
\ee
with $\lambda$ a dimensionless coupling constant.
Assuming minimal kinetic terms for all fields, and normalising the
charges of the $\phi^\pm$ superfields to unity, the $D$-term scalar
potential reads
\be
V_D = \frac{g^2}{2}\left(|\phi^+|^2 - |\phi^-|^2 -
\xi \right)^2
\ee
(we use the same notation for the superfields and their scalar
components).  A special choice of parameters is  obtained 
in the Bogomolnyi limit when $\lambda=\sqrt{2} g$, which is
appropriate for string motivated models of inflation~\cite{D3D7}.
Inflation takes place for $\phi \neq 0$ and $\phi^\pm =0$, in
which case the scalar potential is a constant
\be V = V_0 \equiv \frac{g^2 \xi^2}{2} \, .
\label{V0susy}
\ee
The inflaton direction is completely flat at tree level.  Radiative
corrections, which are non-zero due to SUSY breaking during inflation,
lift the potential and give the required slope for successful
slow-roll inflation~\cite{Dterm}.  Inflation ends when $\phi$ reaches
the critical value $\phi_c=\sqrt{2\xi}(g/\lambda)$ when one of the mass
eigenstates of the charged fields
\be
m^2_\pm = \frac12 \lambda^2 |\phi|^2 \pm g^2 \xi
\label{mpm_susy}
\ee
becomes tachyonic, and the fields roll towards the SUSY preserving
minimum at $\phi^- = \phi =0$ and $\phi^+ = \sqrt{\xi}$. Cosmic strings
form during this symmetry breaking phase transition.

The above model can be realised within a SUSY description of the D3/D7
system of string theory~\cite{D3D7}. Here the inflaton is the
interbrane distance, and the waterfall fields correspond to interbrane
open string degrees of freedom.

It is not straightforward to generalise $D$-term inflation to
local supersymmetry.  The problem is that the FI-term can only
exist if it is charged under a gauged $U(1)_R$ symmetry.
This requires the full superpotential (not just the
inflationary sector, but also the moduli and MSSM sectors) to be
charged~\cite{freedman,pierre}. Needless to say, this complicates
model building enormously. To get around this problem, we will use
an effective FI-term obtained from the VEVs of some other fields
appearing in the $U(1)_1$ $D$-term potential.  There is then no
need for an R-symmetry, and we can more easily include other
sectors in the theory. We will discuss this point in more detail
in section \ref{sec:combo}.

\subsection{Moduli stabilisation with an uplifting $D$-term}
\label{sec:mod}

To stabilise the moduli fields we will use the mechanism
of~\cite{BKQ,ana}. In this set-up background fluxes stabilise all the
complex-structure moduli of a Calabi-Yau compactification in the
context of IIB string theory. One K\"ahler modulus remains to be
stabilised, the volume modulus $T$. This is done with a
non-perturbative superpotential. The model also includes an
`uplifting' $D$-term potential, associated with an anomalous Abelian
symmetry, $U(1)_2$. This term is needed to give a potential whose
minimum is a dS or Minkowski vacuum (which breaks SUSY). If the theory
were uncharged, and the $D$-term were missing, the minimum would
instead be a SUSY preserving AdS vacuum.

This set-up can be described in $N=1$ SUGRA language as follows. The
moduli sector contains the volume modulus $T$ and (at least) one
additional chiral matter superfield $\chi$, charged under
the $U(1)_2$ symmetry.  The chiral field is introduced to
allow for a non-trivial gauge invariant stabilisation superpotential
of the form~\cite{ana}
\be
W_{\rm mod}(T,\chi) = W_0 + \frac{A e^{-a T}}{\chi^b} \, .
\label{Wmod}
\ee
This contains a constant term $W_0$ from integrating out the
complex-structure moduli, and a non-perturbative term from, for
example, gaugino condensation. The parameters $A,a,b >0$ are positive
definite constants. If we define the field behaviour under a $U(1)_a$
gauge transformation via $\delta \phi = \eta_a^\phi \alpha$ where
$\alpha$ is the infinitesimal gauge parameter, the superpotential is
gauge invariant if we choose $\eta_2^T = i\gs^{(2)}/2$ and
$\eta_2^\chi = -ia \gs^{(2)}/(2b)\chi$ with $\gs^{(2)}$ the
Green-Schwarz parameter of the anomalous $U(1)_2$.

In contrast to \cite{ana}, we will use a K\"ahler potential with
the no-scale property
\be
K_{\rm mod} = -3\ln(T+\bar T -|\chi|^2 + \gs^{(2)} \mathcal{V}_2)\, .
\ee
The term proportional to $\mathcal{V}_2$, the gauge superfield
corresponding to
$U(1)_{2}$, is introduced to make the K\"ahler potential manifestly
gauge invariant.  With this choice the $F$-term and $D$-term potentials
are
\be
V^F_{\rm mod} = e^{K_{\rm mod}}
(K_{\rm mod}^{I \bar J}D_I W_{\rm mod} D_{\bar J} \bar W_{\rm mod}
-3 |W_{\rm mod}|^2) \, ,
\label{VFmod}
\ee
\be
V_{\rm mod}^D \equiv V^{(2)}_{D} = \frac{1}{2 \Re f_2(T)}
\left(i\sum_{I=T,\chi} \eta_2^I K_I\right)^2
 =\frac{[3\gs^{(2)}(1+(a/b) |\chi|^2)]^2}{8X^2 \Re f_2(T)} \, ,
\label{VDmod}
\ee
where $I=T,\chi$, and $X= {\rm e}^{-K/3}= 2 T_{\rm R} - |\chi|^2$ with
$T_{\rm R} = \Re T$.  Note that a Minkowski minimum of this theory can
only exist if the uplifting $D$-term is positive definite, which is
only true if $a/b>0$. The no-go theorem that $D$-terms cannot be
used for uplifting due to the SUSY relation $D^a \propto \eta^I_a
W^{-1} D_I W$~\cite{nilles,alwis} is evaded thanks to the
non-analytic form of the superpotential~\eref{Wmod}.

Apart from the presence of $\chi$ the above stabilisation potential is
qualitatively similar to the KKLT scenario~\cite{KKLT}. In KKLT the
uplifting of an AdS $F$-term vacuum is achieved by adding extra anti-D3
branes to the model, which introduces an additional SUSY breaking term in
the potential. In contrast, the scenario outlined above is completely
described by an effective SUGRA theory, and there is no need to
introduce non-supersymmetric terms.

\subsubsection{Parameter estimates}
\label{ssec:Wana}

As we will show later, the coupling of the moduli fields will alter
the dynamics of inflation. To get some idea of how significant these
changes will be, it is useful to have some estimates for the
parameters in the stabilisation sector (\ref{Wmod}--\ref{VDmod}).
Defining the gauge kinetic function for $U(1)_2$ as $f_2(T) = k_2
T/(2\pi)$ with $k_2$ an $\Or(1)$ positive model dependent
constant, consistent anomaly cancellation via the Green-Schwarz
mechanism~\cite{GS} can be used to fix the parameter combination $E =
(3/4) \gs^{(2)} \sqrt{\pi/k_2}$.  Assuming the non-perturbative
potential is generated by gaugino condensation of an $SU(N)$ group on
D7 branes, we find\footnote{These parameter choices follow from
equations (5.6) and (5.7) of \cite{ana}, with $N_f=1$, $k_N=1$,
and $\bar q = q$.}

\be
A = 2^{1/(N-1)} (N-1) \, , \qquad
a = 2\pi b = \frac{4 \pi}{N-1} \, , \qquad
E^2 = \frac{27}{32 \pi N} \, .
\label{Wpar}
\ee
The parameter $W_0$ in \eref{Wmod} is fixed by requiring the
minimum of $V_{\rm mod}= V^F_{\rm mod} + V^D_{\rm mod}$ to be Minkowski.
For example, for $N=15$ we find numerically
\be
T_{\rm R} \approx 6.93 \, , \qquad \chi \approx 0.094 \, , \qquad
W_0 \approx 0.302 \ .
\ee
This differs slightly from the corresponding result in \cite{ana},
which can be attributed to our different choice of K\"ahler potential. The
gravitino mass is $m_{3/2} = e^{K_{\rm mod}/2}W_{\rm mod} \approx
5.1 \times 10^{-3}$. We note that this is quite large and will not
be detected in accelerator experiments. This is a generic feature
of these models. Note that the derivation of the above
superpotential assumes that the KK-particles (with masses of order
$T_{\rm R}^{-1/4}$) can be integrated out. Moreover, the
supergravity approximation must also be valid, i.e.\ the
compactification radius must be larger than the string scale,
resulting in  $T_{\rm R}> 1$.  For the above parameters, this
holds for $N \geq 4$.

\section{Combining inflation and moduli stabilisation}
\label{sec:combo}

In this section we discuss how to combine the inflation and moduli
sectors into a working model.  An important feature of our moduli
stabilisation potential is that the finite $T_{\rm R}$ minimum is
separated by a finite energy barrier from a minimum at $T_{\rm R} \to
\infty$.  For the moduli fields to remain stabilised during inflation,
stabilisation has to take place at a higher scale than inflation.  On
the other hand, the moduli fields should not alter the successful predictions
of $D$-term inflation too much.

For the superpotential of our full theory we take the simplest
possibility:
\be
W = W_{\rm inf}(\phi,\phi^+,\phi^-) + W_{\rm mod}(T,\chi) \, .
\label{inflasec}
\ee
In addition we have to combine the symmetries and K\"ahler potentials
of the two sectors; we will motivate our choices below.

\subsection{Generating a Fayet-Iliopoulos term}
\label{sec:FI}

We now return to the problem of obtaining an effective $\xi$ for the
inflation sector. It is not an easy task to find a consistent SUGRA
theory in which a non-zero Fayet-Iliopoulos term arises from the VEV
of some field~\cite{elvang}.  However, in some sense this problem is
already solved by the moduli sector. An unusual feature of the moduli
stabilisation mechanism that we are using is that its vacuum state has
a non-zero $D$-term. By making $T$ and $\chi$ charged under the
inflation $U(1)_1$ we can make this non-zero quantity appear in the
inflation $D$-term as well, where it can play the role of $\xi$. This
is the setup we will study in this paper.

The symmetry of our model is therefore $U(1)_1 \otimes U(1)_2$, with
both $U(1)$s anomalous. Alternatively, the theory could be rewritten
in terms of two other $U(1)$ symmetries, only one of which is
anomalous (although we find it convenient to not do this). The anomaly
is cancelled via a Green-Schwarz mechanism as in~\cite{ana}.
The fields charged under the inflation $U(1)_1$ are the
waterfall fields $\phi^\pm$ as well as the moduli sector fields
$T,\chi$, with $\eta_1^T = i\gs^{(1)}/2$, $\eta_1^\chi = -ia
\gs^{(1)}/(2b)\chi$, $\eta_1^{\phi \pm} = \pm i\phi^\pm$ and
$\eta_1^\phi=0$. Only the moduli sector fields are charged under
$U(1)_2$, with charges given in section~\ref{sec:mod}.  With these
assignments the superpotential is invariant under both symmetries.
The moduli sector fields are stabilised at finite $T$ and $\chi$,
their VEVs generating an effective FI-term in $V^{(1)}_D$ and also a
constant, uplifting $V^{(2)}_D$. The precise form of the $D$-terms
depends on the K\"ahler potential. If we use~\eref{noscale}
below, then the inflation $D$-term is
\be \fl
V^{(1)}_{D} = \frac{1}{2 \Re f_1(T)}
\left(i\sum_j \eta_1^j K_j\right)^2
=\frac{g^2(T)}{2X^2}\left[|\phi^+|^2 - |\phi^-|^2 -
\frac{3}{2}\gs^{(1)} \left(1 + \frac{a}{b}|\chi|^2\right)
\right]^2 \, ,
\label{VD1}
\ee
with $X = {\rm e}^{-K/3}$, which is equal to $2T_{\rm R}- |\chi|^2$
during inflation, and $g^2 = 1/[\Re f_1(T)] \propto 1/T_{\rm R}$ is
the effective gauge coupling. The effective Fayet-Iliopoulos term in the above
expression is given by
\bea
\frac{3}{2}\gs^{(1)} \left(1 + \frac{a}{b}|\chi|^2\right)
= X i\left(\eta_1^T K_T + \eta_1^\chi K_\chi\right)
= X \frac{\gs^{(1)}}{\gs^{(2)}} \sqrt{2 V^D_{\rm mod} \Re f_2(T)}\, .
\eea
However as it stands, this expression does not quite provide a suitable $\xi$
for our SUGRA extension of SUSY $D$-term inflation. This is partly due
to our use of non-canonically normalised fields. If we take
$\phi^\pm=0$, as is the case during inflation, we want the above
potential term~\eref{VD1} to be equal to the inflation vacuum
energy~\eref{V0susy}. We therefore define the appropriately scaled
\be
\xi = \frac{3}{2}\gs^{(1)}
\frac{1 +(a/b)|\chi|^2}{2T_{\rm R} - |\chi|^2}\, ,
\label{FI}
\ee
which will give $V_0 =g^2 \xi^2 /2$ during inflation. We will use this
expression for $\xi$ throughout the paper.

Hence we see that the positive definite uplifting $D$-term of the moduli
stabilisation sector can be used to give a non-trivial
Fayet-Iliopoulos term. We see that for this approach to work the
moduli field $T$ must be charged, and so the above setup would not
work with the KKLT scenario. In this case it may still be possible to
obtain a non-zero $\xi$ using some additional field. In this paper we
will consider a minimal setup in which the source of the FI-term is
$T$ (and $\chi$).

\subsection{Choice of K\"ahler potential}
\label{ssec:K}

The simplest choice of K\"ahler potential for the inflationary sector
is the minimal
\be K^{\rm (min)}_\mathrm{inf} = |\phi|^2 + |\phi^+|^2 + |\phi^-|^2 \, .
\ee
This could simply be added to the moduli sector K\"ahler potential, so $K =
K_{\rm mod} + K^{\rm (min)}_\mathrm{inf}$. During inflation the full
potential simplifies since $\phi^\pm=0$, and
\be
V = e^{|\phi|^2} (V^F_{\rm mod} + m_{3/2}^2 |\phi|^2) +
V^D_{\rm mod} + V_0 \, .
\label{VKmin}
\ee
The gravitino mass is defined by $m_{3/2} = e^{K_{\rm mod}/2} W_{\rm mod}$,
and $V_0 \equiv V^{(1)}_D = g^2 \xi^2/2$ during inflation. For
simplicity we will take $W_{\rm mod}$ to be real throughout this section.

Note that typically
$m_{3/2} \gtrsim g \xi$ is needed for the moduli to remain stabilised during
inflation.  We see from \eref{VKmin} that the presence of the moduli
sector gives a large mass to  the inflaton. As a result the second
slow-roll parameter $\eta = V''/V$ is large, prohibiting slow-roll
inflation.  This is the infamous $\eta$-problem of SUGRA
inflation~\cite{eta}.  Normally absent in SUGRA $D$-term inflation,
the $\eta$-problem has reappeared through the coupling to the moduli sector.

The $\eta$-problem may be cured by fine tuning parameters, but a more
elegant solution is to introduce a shift symmetry for the inflaton
field in the K\"ahler potential
\be
K_\mathrm{inf} = \frac{|\phi -\bar \phi|^2}{2}
+ |\phi^+|^2 + |\phi^-|^2 \, .
\label{Kinf}
\ee
Such a shift symmetry arises naturally in models of brane inflation as
a consequence of the translational invariance of the brane system
(see e.g. the discussions in~\cite{braxshift} and references therein).
With the inflaton the normalised real part of $\phi$, we now find
$V=(V^F_{\rm mod} + V^D_{\rm mod}) + V_0$ during inflation; the
inflaton potential is again flat at tree level.

Taking $K = K_{\rm mod} + K_\mathrm{inf}$ we can calculate the mass
spectrum during inflation.  The normalised imaginary part of $\phi$
has mass $2 m_*^2$ where
\be
m_*^2 = 2m_{3/2}^2 + V^F_{\rm mod} \, .
\ee
We want this mass to be large and positive during inflation so that
$\Im \phi$ is stabilised.  However, typically $V^F_{\rm mod} \sim
-3m_{3/2}^2$, and so $\Im \phi$ is tachyonic. This can be avoided by
tuning the parameters in the moduli sector. For the
parameters~\eref{Wpar} $m_*^2 >0$ only when $N < 7$.
The mass eigenstates of the waterfall fields also get contributions
from the moduli sector:\footnote{$\phi = \Re \phi$ since $\Im \phi =0$
during inflation.}
\be
m_\pm^2 = \frac{\lambda^2}{X^3} \phi^2 + m_*^2 - m_{3/2}^2
\pm \left\{\left[2-\frac{X W_T}{W_{\rm mod}}\right]^2
m_{3/2}^2 \frac{\lambda^2}{X^3} \phi^2 + g^4 \xi^2\right\}^{1/2}
\label{mpm1}
\ee
with $W_T = \partial W_{\rm mod}/ \partial T$ and as before
$X = {\rm e}^{-K/3}$, which is $2T_{\rm R} - |\chi|^2$ during
inflation. This poses a more
serious problem.  For the moduli to be stabilised during inflation the
stabilisation scale has to be larger than the inflationary scale. As a
consequence the moduli contribution $m_*^2 - m_{3/2}^2$ in \eref{mpm1}
dominates the mass.  This term is generically tachyonic, the
fine-tuning needed to make it positive is more severe than for $m_*^2$.
And even if it can be tuned positive, it is generically large and
$m_-^2$ never becomes tachyonic.  There is no phase transition to end
inflation, hence no graceful exit.

Some of the moduli sector contributions can be cancelled by
introducing an additional shift symmetry for the waterfall
fields~\cite{Fterm}
\be
K^{\rm (shift)}_\mathrm{inf} = \frac12 |\phi
-\bar \phi|^2 + |\phi^+ -\bar \phi^-|^2 \, .
\ee
Note that this additional shift symmetry can only be imposed if
$\phi^\pm$ have equal and opposite charges. Had we used a constant
FI-term as opposed to an effective one, the charges of the waterfall
fields would have been shifted~\cite{pierre}, and we could not have
used this form of the K\"ahler potential. The squared mass of $\Im \phi$ is
unchanged, and tuning is still needed to make it positive.  The masses of
the waterfall fields are now
\be \fl
m_\pm^2 = \frac{\lambda^2}{X^3} \phi^2 + m_*^2 - 2  m_{3/2}
\frac{\lambda}{X^{3/2}} \phi  \pm\left\{ \left[m_*^2 - \left(2-\frac{X
W_T}{W_{\rm mod}}\right) m_{3/2} \frac{\lambda}{X^{3/2}} \phi\right]^2
\! + g^4 \xi^2 \right\}^{1/2}  \! ,
\ee
hence
\be
m_-^2 = \frac{\lambda^2}{X^3} \phi^2
- \frac{\lambda |W_T|}{X^2} \phi -\frac{\xi^2 g^4}{2 m_*}
 +\Or\left(\frac{g^8\xi^4}{m_*^3},\frac{g^4\xi^2\phi}{m_*}\right)
\, ,
\ee
It is now possible for the end of inflation to occur when $\phi$ is
small. However we see that the value of $\phi_c$ has little resemblance
to that in the usual SUSY $D$-term inflation model.  This contrasts
with $F$-term hybrid inflation, where the above shift symmetry does give
a conventional value for $\phi_c$ (with $\Or(W_T)$
corrections)~\cite{Fterm}.  We also discard this form for the
K\"ahler potential.

The reason that the inflation sector fields are receiving large (and
typically negative) contributions to their masses from the moduli
sector, is that they couple to $V^F_{\rm mod}$ (which is large and
negative), but do not couple to the lifting term $V^D_{\rm mod}$
(large and positive), which has been chosen to cancel the $F$-term. By
coupling the inflationary fields to the whole moduli stabilisation
potential, the above problem can be avoided. This can be achieved by
taking the full K\"ahler potential to be no-scale
\be
K= -3\ln\left(T +\bar T - |\chi|^2 - K_\mathrm{inf}/3
 + \gs^{(1)} \mathcal{V}_1 + \gs^{(2)} \mathcal{V}_2\right) \,
\label{noscale}
\ee
with $K_{\rm inf}$ as given in~\eref{Kinf}, i.e., with a shift
symmetry for the inflaton field to solve the $\eta$-problem. Note
that $\phi$ and $\phi^\pm$ are no longer canonically normalised.  With
this choice
\be m_*^2 = \frac{|W_T|^2}{9X} +
\frac{g^2 \xi^2}{3}
\label{mphi}
\ee
is positive definite, and so $\Im \phi$ is never tachyonic. The
masses of the waterfall fields are
\bea
m^2_\pm &=& \frac{2 V_0}{3} +\frac{\lambda^2}{X} \phi^2
 \pm g^2 \xi \left[1 +
\frac{\lambda^2 W_T^2}{9(g^2 \xi)^2 X^2} \phi^2 \right]^{1/2}
\nonumber \\
&=& g^2 \xi
\left(\frac{\xi}{3}+ \frac{\phi^2}{\phi_1^2}
\pm \sqrt{1 + \frac{2}{3}\alpha \xi \frac{\phi^2}{\phi_1^2}}\right)
\label{mpm}
\eea
with
\be
V_0 = \frac12 g^2 \xi^2\, , \qquad
\phi_1^2 = \frac{g^2 \xi X}{\lambda^2} \, , \qquad
\alpha = \frac{W_T^2}{6 g^2 \xi^2 X} \, .
\label{phi1}
\ee
The quantity $\alpha$ is related to the ratio of the moduli and
inflation energy scales. It will play an important role in this paper,
as its value determines how much the observational predictions of our
SUGRA model deviate from those of the corresponding SUSY $D$-term
inflation model.

In the absence of the moduli corrections (the terms proportional to
$V_0$ and $W_T$), and after canonically normalising the inflaton
field, this reduces to the standard SUSY $D$-term inflation results
\eref{mpm_susy}. In that case the phase transition triggering the end
of inflation occurs when $\phi$ reaches $\phi_c = \phi_1$; the moduli
corrections change this to
\be
\phi^2_c = \phi_1^2 \left( \sqrt{1+\frac{\xi^2}{9} \alpha (\alpha-2)}
+\frac{\xi}{3}(\alpha-1)\right) \, .
\label{phiend}
\ee
Numerically, we find that  for the parameter choices~\eref{Wpar}, $W_T
\approx  W/T$ and  $X \approx  2T$  for a  wide range  of $N$.  Hence,
barring fine tuning of the moduli potential, we need
\be
\alpha \approx \frac{m_{3/2}^2}{3V_0} > 1\, .
\label{alpha2}
\ee
for the moduli fields to remain stabilised during inflation.
Numerically, we typically find  $\alpha \gg 1$.  For example
$\alpha \approx 8 \times 10^{4}\, (10^{-10}/V_0)$
for $N=15$.

\section{Effective Potential}
\label{sec:effepo}
\subsection{Tree level potential}

The combined theory for the moduli and inflation sectors that we will
be studying uses the K\"ahler potential~\eref{noscale}
\be
\fl K= -3\ln\left\{T +\bar T - |\chi|^2
+ \gs^{(1)} \mathcal{V}_1 + \gs^{(2)} \mathcal{V}_2
- \frac{1}{3}\left(
\frac{|\phi -\bar \phi|^2}{2}
+ |\phi^+|^2 + |\phi^-|^2\right)\right\} \, ,
\ee
and the charge
assignments for the fields are discussed in section~\ref{sec:FI}.
The superpotential is given by \eref{inflasec}, with the two
contributions to it being \eref{Winf} and \eref{Wmod}. Combining all
this produces the scalar potential
\be
V = V_{\rm mod} + V_{\rm inf} + V_{\rm mix}
\ee
with
\bea \fl
V_\mathrm{mod} =
\frac{1}{X^2} \Bigg(
\frac{1}{3}\left[2T_{\rm R} |W_T|^2 + |W_\chi|^2
 +2\Re\{(\chi W_\chi-3 W_{\rm mod}) \bar W_{\bar T}\}\right]
\nonumber \\ \fl \hspace{3in} {}
+\frac{[3\gs^{(2)}(1+(a/b) |\chi|^2)]^2}{8\Re f_2(T)}
\Bigg)\,,
\eea
\be \fl
V_{\rm inf} =
\frac{1}{X^2} \Biggl( \lambda^2
\left[|\phi^- \phi^+|^2 + |\phi^- \phi|^2  +|\phi^+ \phi|^2 \right]
+ \frac{g^2}{2}\left[|\phi^+|^2 - |\phi^-|^2
- (2T_{\rm R} - |\chi|^2)\xi\right]^2\Biggr)\, ,
\ee
\be \fl
V_{\rm mix} =
\frac{1}{X^2} \left(-\frac{(\phi-\bar \phi)^2}{18} |W_T|^2
-\frac{2\lambda}{3}\Re\left[\phi W_T \bar\phi^+ \bar \phi^- \right]
\right)\,.
\ee
Here $X = {\rm e}^{-K/3} = 2T_{\rm R} - |\chi|^2 - K_{\rm inf}/3$, and
the FI-term is~\eref{FI}
\be
\xi = \frac{3}{2}\gs^{(1)}
\frac{1 +(a/b)|\chi|^2}{2T_{\rm R} - |\chi|^2}\,.
\ee
During inflation $\phi^\pm=0$ and $\Im \phi=0$, so the above
expression reduces to
\be
V = \frac{g^2\xi^2 }{2} + V_\mathrm{mod}\, .
\ee
Significantly, it is completely
independent of the inflaton $\Re \phi$, and (assuming the moduli
are stabilised) is a constant. The moduli sector does not give any
tree-level contributions to the inflaton slope. Hence we appear to be
in a similar situation to conventional $D$-term inflation. This is in
sharp contrast to the $F$-term inflation
model discussed in~\cite{Fterm}.

After inflation $\phi=0$, $\phi^-=0$ and
$\phi^+= \sqrt{(2T_{\rm R} -  |\chi|^2)\xi}$.
The VEV of $\phi^+$ is different to that in the corresponding SUSY model,
although this is partially due to the different normalisation of the
fields. For a canonically normalised field, it would be
$\sqrt{\xi/(1- \xi/3)}$, which is closer to the SUSY value $\sqrt{\xi}$. The
potential reduces to $V= V_\mathrm{mod}$, and now
$X = (2T_{\rm R} - |\chi|^2)(1- \xi/3)$.

Note that after canonically normalising the fields $\phi^I \to
\phi^I/\sqrt{X}$, the inflationary potential $V_{\rm inf}$ is
almost identical to that of standard $D$-term inflation.  From now on we will
use the canonically normalised inflaton field
\be
\vp = \sqrt{\frac{2}{X}} \Re \phi \, .
\label{varphi}
\ee

\subsection{Loop corrections}

The one-loop corrections to the effective potential for a general
theory, with cutoff scale $Q$, are~\cite{coleman,loop}
\be
V_\mathrm{loop} = \frac{1}{32 \pi^2} \mathrm{Str} M^2 Q^2
+\frac{1}{64\pi^2} \mathrm{Str} M^4 \ln \frac{M^2}{Q^2} \, ,
\ee
where the supertrace is defined by
$\mathrm{Str} f(M)  = f(M_{\rm (boson)}) - f(M_{\rm (fermion)})$.
We will compute the contributions to $V_\mathrm{loop}$
during inflation. The only $\vp$-dependent contributions come from
the masses of the waterfall fields given in \eref{mpm} and of their
superpartners $\tilde{m}_\pm = \lambda \vp /\sqrt{2}$
%
\be
\left( V_{\rm loop} \right)_{\phi_\pm} = \frac{g^2 \xi^2 Q^2}{24 \pi^2}
+\frac{g^4\xi^2}{32 \pi^2} U(\vp/\vp_1)
\label{Vmpm}
\ee
with $\vp_1^2 =2 g^2\xi/\lambda^2$, see \eref{phi1}, and
\bea
U(x) &=&
\left(x^2+\xi/3 - \sqrt{1+ 2\alpha x^2 \xi/3}\right)^2
\ln\left(x^2+\xi/3 - \sqrt{1+ 2\alpha x^2 \xi/3}\right)
\nonumber \\ && {}
+\left(x^2+\xi/3 + \sqrt{1+ 2\alpha x^2 \xi/3}\right)^2
\ln\left(x^2+\xi/3 + \sqrt{1+ 2\alpha x^2 \xi/3}\right)
\nonumber \\ && {}
- 2x^4 \ln x^2
+ \frac{4 \xi}{3} (1+\alpha) x^2 \ln \frac{g^2 \xi}{Q^2}
+2 \left(1+\frac{\xi^2}{9}\right)\ln\frac{g^2 \xi}{Q^2}
\, .
\eea
Note that in contrast to the corresponding SUSY model, inflation does
not end at $x=1$. However if $\xi$ and $\alpha \xi$ are small, it will
end close to $x=1$.

Setting $\alpha=0$ and taking the limit of small $\xi$, the moduli corrections
switch off. To leading order in $\xi$, the form of the potential
reduces to the usual expression for $D$-term hybrid inflation
\be
U(x) = (x^2 - 1)^2 \ln(1 -x^{-2}) + (x^2+1)^2 \ln(1 + x^{-2})
+ 2\ln \frac{g^2 \xi}{Q^2}x^2 \, .
\label{Ususy}
\ee
The contribution from the masses~\eref{mpm} provides the slope in the
potential, enabling slow-roll inflation.

The contribution of all other fields to $V_{\rm loop}$ gives (assuming
that the moduli fields are fixed) just constant contributions to
the potential, which can be absorbed into the FI-term and $V_{\rm mod}$.
For example, the inflaton is massless, the mass of $\Im \phi$ is $2
m_*^2$ [see \eref{mphi}], and the inflatino mass is 
$\tilde{m}_\phi = -W_T/(3\sqrt{X})$. This gives a contribution
\be \fl
\left(V_{\rm loop}\right)_{\phi} =
\frac{g^2 \xi^2Q^2 }{48 \pi^2}  +
\frac{g^4 \xi^4 }{288 \pi^2}
\Bigg[ (2\alpha+1)^2 \ln  (2\alpha+1) - (2\alpha)^2 \ln  (2\alpha)
+(1+4\alpha)\ln \Big(\frac{g^2\xi^2}{3 Q^2} \Big)
\Bigg] \, .
\ee
Similarly, the moduli and gauge fields give constant contributions to
the potential.  As a result, the $\vp$ dependence of the
one-loop corrections to the potential is not affected by the other
fields, remaining a sole function of $U(\vp/\vp_1)$.  The
$\xi$-dependent constants in $V_{\rm loop}$ are absent after
inflation;  during inflation  they rescale the
FI-term. The $\xi$-independent constants can be absorbed in
$V_{\rm mod}$. The full inflationary potential must read
\be
V = \tilde V_0 +
\frac{g^4 \xi^2}{32\pi^2} U(\vp/\vp_1)
+ \tilde V_{\rm mod}
\label{Vinftilde}
\ee
with the tilde denoting the  rescaled quantities.
\be
\tilde V_0 = \frac{g^2 \xi^2}{2}
\left[1+ \frac{Q^2}{8\pi^2}
+ \Or(\alpha^2 g^2 \xi^2) \right]
\label{V0tilde}
\ee
is the  vacuum energy during inflation. The loop corrections
to the FI-term are subdominant, and perturbation theory is valid, for
$g \xi \alpha \lesssim 1$. As we will see in subsection~\ref{ssec:dV},
$\alpha \xi \lesssim 1$ must be small for inflation to work, so this is
indeed the case. For notational convenience we will drop the tildes
from now on.

\subsection{Moduli variation}

It is usually assumed that moduli fields are stabilised before
inflation begins, and that they remain fixed during inflation.
If the energy scale of the stabilisation potential
is much higher than that of inflation, the effects of inflation will
not significantly alter the values of the moduli fields, and this
seems a reasonable assumption. However, care should be taken.  Since
the inflaton potential is very flat, even tiny changes in the moduli
sector may affect it significantly. Indeed, this was shown to be the
case for $F$-term hybrid inflation model discussed in~\cite{Fterm}. In
this paper we will be assuming that the moduli do not vary during
inflation. We will now  determine when this really is a
reasonable assumption for our model.

The inflaton potential depends on the moduli fields $T$ and $\chi$ via
the effective parameters $g$, $\xi$ and $\alpha$.  As a consequence
their minimum will shift slightly as the inflaton field rolls down its
potential, and inflation takes place. Let us estimate the
effects of this moduli variation on the inflaton potential.
During inflation the full potential is
\bea \fl
V(\phi,Y^c) &=  V_\mathrm{mod}(Y^c) + V_{\rm inf}(\phi,Y^c)
\nonumber \\ \fl
&= V_{\rm inf}(\phi,Y^c_0)
+\frac{1}{2}  M_{ab} \delta Y^a \delta Y^b
+ V_{{\rm inf},a}(\phi,Y^c_0) \delta Y^a
+\Or\left(\delta Y^3,\delta Y^2 V_{\rm inf}\right)
\hspace*{0.3in}
\eea
where $Y^a = \{T_{\rm R}, |\chi| \}$.  $V_\mathrm{mod}$ and $V_{\rm
inf}$ include the 1-loop corrections. In the second line we Taylor
expanded around $Y_0^c$, with $Y_0^c$ being the minimum of $V_{\rm mod}$. We
further introduced the notation $M_{ab} = V_{\mathrm{mod},ab}
(Y^c_0)$. The potential is minimised by $\delta Y^a = -M^{-1}_{ab}
V_{{\rm inf},a}$.  Substituting this back in gives
\be
V(\phi) \approx V_{\rm inf}
- \frac{1}{2} V_{{\rm inf},a} M^{-1}_{ab} V_{{\rm inf},b} \, .
\ee
For the parameter choices~\eref{Wpar}  we expect that
\be
V_{{\rm inf},a} \sim \frac{V_{\rm inf}}{T_{\rm R}} \, \qquad
M_{ab} \sim \frac{m_{3/2}^2}{T_{\rm R}^{2}} \, .
\ee
Hence, the
moduli variation $\delta Y^a \sim T_{\rm R} V_{\rm inf} /
m_{3/2}^2$ is indeed small. The correction to the inflationary
potential is of the order
\be
\delta V_{\rm inf} \sim
\frac{V_{\rm inf}^2}{m_{3/2}^2} \sim \frac{
V_{\rm inf}}{\alpha}
\ee
which is small compared to $V_\mathrm{inf}$ and can safely be neglected
when $\alpha \gg 1$.  Thus for the parameter choices such as
\eref{Wpar}, if the moduli scale is much larger than the inflationary
scale and $\alpha$ is large, it is a good assumption to take the moduli fixed.

\subsection{Potential during inflation}
\label{ssec:dV}

\begin{figure}
\centerline{\includegraphics[width=7.5cm]{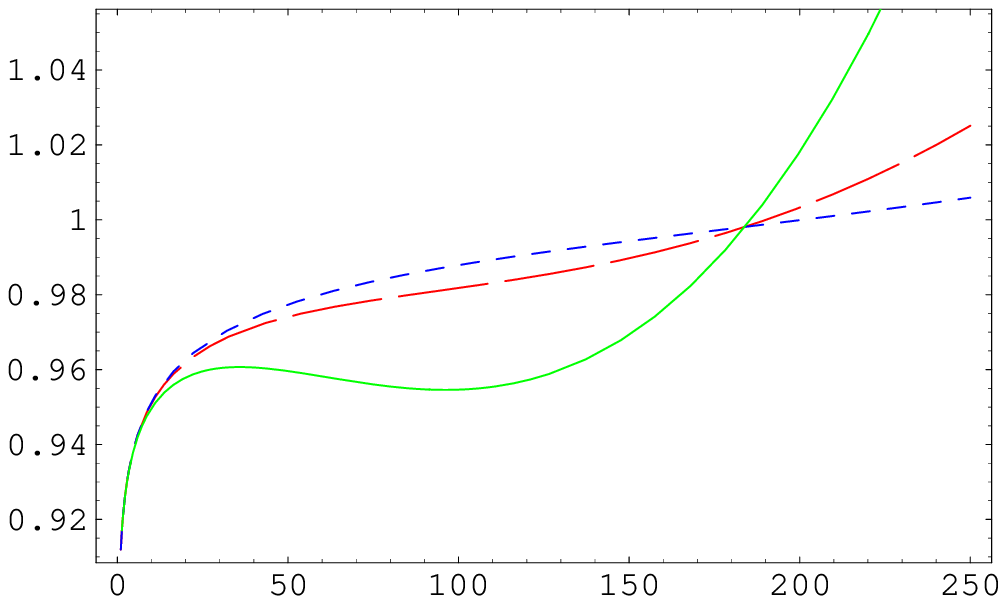} 
\includegraphics[width=7.5cm]{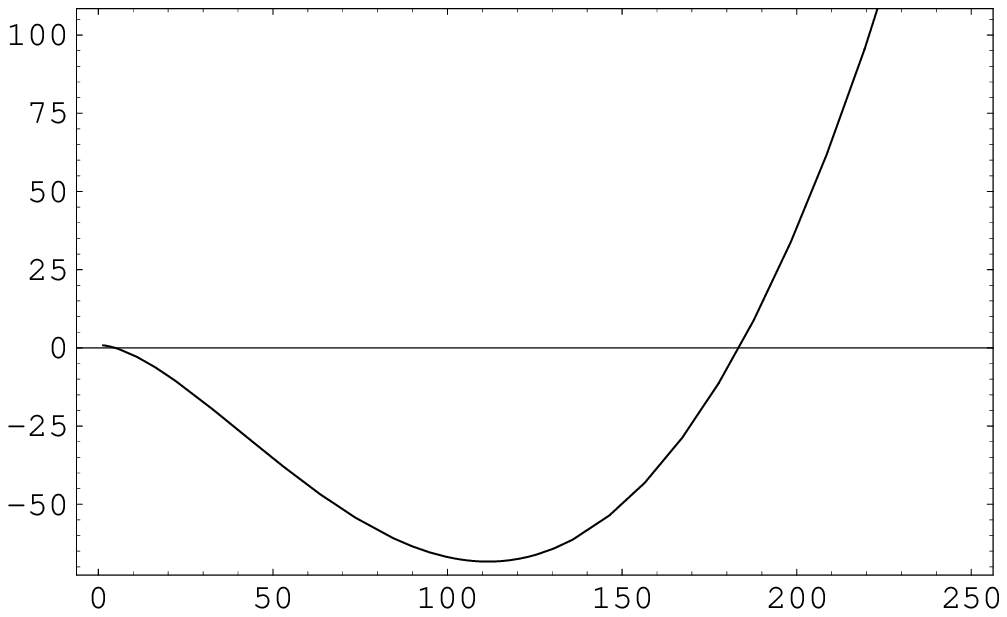}}

\vspace{-0.6in} \hspace{2.6in}
(a) \hspace{2.7in} (b) \vspace{0.4in}

\caption{
$V/V_0$ as a function of $x=\vp/\vp_1$.  In figure (a) the curves
correspond to parameter choices with $\mathcal{B}>1$ (upper, green),
$\mathcal{B}=1$ (middle, red), and $\mathcal{B} <1$ (lower, blue).
Figure (b) is for $\alpha \xi = 1$.}
\label{fig:V}
\end{figure}

Before delving into the details of the inflationary dynamics, we will
discuss some general properties of the inflaton potential
\eref{Vinftilde}.
At $\vp =\vp_c$ one of the waterfall fields becomes tachyonic,
triggering the symmetry breaking phase transition ending
inflation\footnote{In large parts of parameter space the slow roll
parameters exceed unity, and inflation is ended, for $\vp > \vp_c$.}.
If $V'(\vp_c) <0$, as in \fref{fig:V}b, the
inflaton will never reach this value.  Instead it will end up in the
wrong vacuum at the minimum of $V_{\rm inf}$ occurring at $\vp \gg \vp_c$. 
In the $\xi \ll 1$ limit
\bea \fl
V'(\vp_c) &= \frac{g^2 \lambda^2 \xi \vp_c}{8\pi^2}
\left\{
\ln 2\sqrt{1+\frac{\alpha^2 \xi^2}{9}}
+ \frac{\alpha \xi}{3} \left[\frac{1}{2}
+ \mathrm{arcsinh}\,\frac{\alpha \xi}{3}
+ \ln \frac{4g^2\xi}{Q^2}\right] + \Or(\xi)\right\}
\nonumber \\ \fl
&= \frac{g^2 \lambda^2 \xi \vp_c}{8\pi^2}
\left\{ \ln 2 + \frac{\alpha \xi}{3} \left[
\ln \frac{g^2\xi}{Q^2} + \frac{1}{2} + 2\ln 2\right]
+ \Or(\xi, \alpha^2\xi^2)\right\} \, .
\label{dVc}
\eea
which becomes negative for larger values of $\alpha \xi$.  Requiring
$V'(\vp_c) > 0$, to ensure that the system ends up in the right vacuum,
therefore gives us an upper bound on $\alpha$. For example if
$g^2 \xi/Q^2 = 10^{-6}$ then $\alpha \xi \lesssim 0.175$ is the
required bound\footnote{In fact \eref{dVc} becomes positive again for
very large $\alpha \xi$; such large values of $\alpha$ are outside the
range of validity of our effective theory, see the discussion below
\eref{V0tilde}.}. Recall that $\alpha$ parametrises the ratio of the
moduli to inflationary scale (\ref{phi1}, \ref{alpha2}), which should be
larger than unity to keep the moduli fields stabilised during
inflation.  Successful inflation with the moduli stabilised, and with the 
inflaton ending up at the critical value $\vp_c$, is possible when 
(in the $\xi \ll 1$ limit)
\be
1 \lesssim \alpha \lesssim \frac{3 \ln 2}{\xi \ln(Q^2/[g^2 \xi])} \, .
\label{alpha3}
\ee
The critical value of $\vp$ at which inflation ends is then
\be
\vp_c = \frac{g \sqrt{2 \xi}}{\lambda} [1+\Or(\alpha \xi)] \, .
\ee
In terms of the gaugino condensation parameters of the $SU(N)$ group
in~\eref{Wpar}, $\alpha$ in the above range requires $24 \leq N \leq 250$, 
for $g^2 =0.1$, $\xi = 10^{-5}$. This is a wide range of
acceptable $N$, though large gauge groups are needed.  For smaller $N$
the combination $\alpha \xi$ is too large.

The inflaton potential features minima and maxima (see~\fref{fig:V}a).  
To better understand their occurrence, we perform a
$\vp \gg \vp_1$ asymptotic expansion. Taking $\xi \ll 1$ (as implied
by the COBE bound~\eref{COBE}), the asymptotic inflationary potential
reduces to
\be \fl
V_{\rm inf} \approx \frac{g^2\xi^2}{2}\left\{ 1
+ \frac{g^2}{8\pi^2}\ln\frac{\lambda^2\vp^2}{2Q^2 e^{-3/2}}
+\frac{\lambda^2 \vp^2}{24\pi^2} \left[ (\alpha +1)
 \ln \frac{\lambda^2 \vp^2}{2Q^2 e^{-3/2}} - 1 \right]
\right\} \, ,
\label{Vinf}
\ee
with the next higher order corrections being $\Or(\vp_1^2/\vp^2)$.
The above limit would not be valid if $\alpha \xi \gg 1$, but we
already know this can not be the case from \eref{alpha3} above.
We see that the slope of the above potential is sensitive to the
choice of cutoff scale $Q$, in contrast to SUSY hybrid
inflation models. The corrections to $n_s$ from the moduli sector
are also strongly dependent on $Q$. Before precise predictions are
extracted from the model,  $Q$ needs to be absorbed into the other
parameters by renormalisation. This  will not be covered
here. We expect the errors to be minimised by taking the renormalisation
scale to be of order $\lambda \vp_*$.

In usual $D$-term hybrid inflation $V$ is monotonic. However because of
the $\vp^2 \ln \vp^2$ correction from the coupling to the
moduli sector, this is no longer true.  To find the extrema, we
calculate the derivative of the above asymptotic potential~\eref{Vinf}
\be
V'_{\rm inf} \approx \frac{g^4 \xi^2}{8\pi^2} \frac{1}{\vp}\left[
1 +  \frac{\vp^2}{\mathcal{B} \vp_x^2}\left(
\ln\frac{\vp^2}{\vp_x^2} - 1 \right) \right] \, ,
\ee
where
\be
\vp_x^2 = \frac{2Q^2}{e^{5/2}\lambda^2}
e^{1/(\alpha+1)}
\, , \qquad \mathcal{B} = \frac{3 g^2}{\lambda^2 (1+\alpha)\vp_x^2} \, .
\label{Bphix}
\ee
Hence $V'=0$ whenever $-\vp^2 \ln [{\vp^2}/({e\vp_x^2})] =
\mathcal{B} \vp_x^2$. If $\mathcal{B}$ is small this equation has
two solutions and the potential has a maximum and a minimum, with
$0 < \vp_{\rm max} < \vp_x < \vp_{\rm min}$.  This is true up
until the critical value $\mathcal{B}=1$, where the two extrema merge
at $\vp = \vp_x$. For larger $\mathcal{B}$ the slope of the
potential is never zero.  This is illustrated in 
\fref{fig:V}a, which shows $V$ for various choices of $\mathcal{B}$.

The value of $\mathcal{B}$ depends on $\alpha$, which is sensitive to
the details of the moduli sector. Since $\alpha \geq 0$ we find an upper
bound of
\be
\mathcal{B} = \frac{3g^2e^{5/2}}{2Q^2 (1+\alpha)e^{1/(1+\alpha)}}
\lesssim 18.3 \, \frac{g^2}{Q^2} \, .
\label{Bmax}
\ee
If the moduli sector has the form described in subsection~\ref{ssec:Wana}, then
we can use the approximate expression~\eref{alpha2} for
$\alpha$. Assuming $\alpha \gg 1$, we obtain
\be
\mathcal{B} \approx 27.4 \, \frac{g^4 \xi^2}{Q^2 m_{3/2}^2} \, ,
\ee
so in this case we usually have $\mathcal{B} \ll 1$.
More generally, we expect $g$ to be small and $\alpha$ to be large,
so the maximum (and minimum) are typically present. For the limiting case
$\mathcal{B} \ll 1$, or equivalently $(1+\alpha) \gg g^2$, we find
\be
\vp^2_{\rm max}
\approx \frac{3 g^2}{\lambda^2 (1+\alpha)\ln (e/\mathcal{B})} \, .
\label{aphimax}
\ee

The presence of the maximum is solely due to the coupling to the
moduli sector. If the initial value of the inflaton is larger that
$\vp_{\rm max}$, it will roll to the minimum of $V_{\rm inf}$ and
inflation will never end. To avoid this we must either have
$\mathcal{B} \geq 1$ or tune the initial conditions for inflation.
This initial value problem will be most severe for models where inflation
needs to take place close to the maximum\footnote{There is a similar initial
value problem in the modulus sector.}.

\section{Slow-Roll Inflation}
\label{sec:inf}

\subsection{Matching with observations}

In the previous section we derived the inflationary potential for our
model, including the one-loop effects. The parameters appearing in
this potential are constrained by the data, since the density
perturbations produced should match the observed ones.

We recall here the important formulae for the density
perturbations. Observable scales leave the horizon $N_* \approx 60$
$e$-folds before the end of inflation, with

\be
N(\vp) \approx \int^{\vp}_{\vp_{\rm end}} \frac{V}{V'} {\rm d} \vp,
\label{N}
\ee
in the slow-roll approximation.
Here $\vp_{\rm end}$ is the value of $\vp$ when inflation ends,
which is either the critical value $\vp_c$ where one of the
waterfall fields becomes tachyonic, or the
value for which one of the slow roll parameters exceeds
unity\footnote{In the limit $\vp_* \gg \vp_{\rm end}$ the integral
\eref{N} is insensitive to the value $\vp_{\rm end}$ and the error made
by setting $\vp_{\rm end} = 0$ is small.}.  The slow roll parameters
are defined as
\be
\eta = \frac{V''}{V} \, , \qquad
\epsilon = \frac{V'^2}{2V^2}
\ee
and $V \approx V_0$ during inflation.  The amplitude of the
perturbations is
\be
\delta_H = \frac{1}{5\sqrt{3}\pi}\frac{V^{3/2}}{V'} \Big|_{\vp = \vp_*}
\label{COBE}
\ee
where $\vp_*$ is the value of the inflaton $N_*$ $e$-folds before the end of
inflation. This should match the COBE normalisation $\delta_H \approx
1.91 \times 10^{-5}$~\cite{Lyth,Cobe}. The spectral index
\be
\label{nsconstr}
n_s = 1 + 2\eta - 6\epsilon \,
\ee
is also constrained by the data. Although the precise observed value
of $n_s$ is sensitive to the choice of prior, it will be in the range
$0.9 \lesssim n_s \lesssim 1$. In hybrid inflation the second
slow-roll parameter $\epsilon$ is usually very small, which implies a
negligible tensor-to-scalar ratio $r = 12.8\epsilon$. In this case,
the WMAP 3-year data prefer smaller values for the spectral index: $n_s =
0.951^{+0.015}_{-0.019}$ \cite{WMAP3}.

The formation of cosmic strings at the end of hybrid inflation gives
additional constraints on the theory, since the string tension $\mu$
is constrained by the data. Strings can contribute a few percent to the
density perturbations, which bounds $G\mu \lesssim 3\times
10^{-7}$~\cite{pogosian,bevis}. The non-observation of irregularities in
pulsar timing experiments bounds the stochastic gravitational wave
background produced by strings, which can be translated into the bound
$G\mu \lesssim 1 \times 10^{-7}$~\cite{Lommen,ShelVil,Vilenkin}. The
mass-per-unit length of a SUSY $D$-term string is directly related to
the Fayet-Iliopoulos term by $\mu = 2 \pi \xi$~\cite{prd}. The
presence of the extra moduli fields will give corrections to this,
which could help satisfy the constraints on $\mu$.

\subsubsection{Small corrections to SUSY hybrid inflation}
\label{ssec:SUSY}

We will begin by determining for which parameter ranges the moduli
corrections are small, and our SUGRA model behaves in a similar way to the
corresponding SUSY model. As is usually done with SUSY hybrid
inflation, we will consider two limiting possibilities for the ratio
$\vp_*/\vp_c$. Recall that $\vp_*$ is the value of $\vp$ when the density
perturbations are produced, and $\vp_c$ is its value when inflation ends.

In the limit that $\vp_* \gg \vp_c$, we can use the asymptotic form of
the inflaton potential~\eref{Vinf}.  As we will see, this is a good
approximation for superpotential couplings which are not excessively
small $\lambda \gtrsim 10^{-5}$. In this limit \eref{N} can be written as
\be
\frac{g^2}{2\pi^2} \frac{dN}{d\vp^2} =
\left[ 1 + \frac{\vp^2}{\mathcal{B}\vp_x^2}
\ln\frac{\vp^2}{e \vp_x^2} \right]^{-1} \, .
\label{Nint}
\ee
The effects of the moduli corrections will be small when
$\vp^2 \ll \mathcal{B} \vp_x^2 = 3g^2/[\lambda^2(1+\alpha)]$.
In this limit, the right hand side of
equation~\eref{Nint} is approximately 1 throughout inflation. Taking
the first order moduli corrections into account, we find
\bea
\frac{g^2N}{2\pi^2}
= \vp^2 +
\frac{\vp^4}{4 \mathcal{B}\vp_x^2}
\left(3-2 \ln \frac{\vp^2}{\vp_x^2}\right)
+\Or\left(\frac{\vp^6}{\mathcal{B}^2 \vp_x^6}\right) \, .
\label{expansion}
\eea
Inverting the above expression, and changing to the usual inflationary
parameters, then gives
\be
\vp_*^2 \approx \frac{g^2 N_*}{2\pi^2}
\left[
1 + \frac{\lambda^2 (1+\alpha) N_*} {12\pi^2}
\left\{2 -\frac{1}{1+\alpha} +\ln \frac{N_* g^2 \lambda^2} {4\pi^2Q^2}
\right\}
\right] \, .
\ee
When $g^2 \lambda^2/Q^2 \lesssim 0.1$, the moduli corrections make the
potential less steep, and reduce the value of $\vp_*$  with respect  to
the standard SUSY result $\vp_*^2 = g^2 N/(2\pi^2)$.  Just as in SUSY
hybrid inflation the first slow roll parameter $\epsilon \ll \eta$ is
small; the spectral index becomes
\be
n_s \approx 1+2\eta \approx 1-\frac{1}{N_*}
+ \frac{\lambda^2 (1+\alpha)}{4\pi^2}
\left\{\frac{11}{3}-\frac{1}{1+\alpha}
+\ln \frac{N_* g^2 \lambda^2}{4\pi^2Q^2}\right\}
\label{nsmod}
\, .
\ee
The SUSY result for the spectral index is $n_s \approx 1+2\eta \approx
1-1/N_* \approx 0.98$.  For $g^2 \lambda^2/Q^2 \gtrsim 0.1$ the
moduli corrections increase this value. In 
the opposite limit $g^2 \lambda^2/Q^2 \lesssim 0.1$, the spectral index is
reduced and can match the best fit WMAP3 value $n_s = 0.95$. However,
this requires the moduli corrections to be substantial, and the
result is sensitive to the details of renormalisation.

The amplitude of the density perturbations should match the COBE
result.  This fixes the FI-term
\be
\xi =  \frac{5\sqrt{3}\delta_H}{2\sqrt{N_*}}\left[ 1
+\frac{\lambda^2 (1+\alpha) N_*}{8 \pi^2}
\left\{\frac{8}{3}-\frac{1}{1+\alpha}
+ \ln \frac{N_* g^2 \lambda^2}{4\pi^2Q^2}\right\}
\right]
\ee
and gives $\xi \sim 10^{-5}$ as usual.

The large-$\vp$ asymptotic expansion that we have used in the above
analysis is only valid if $\vp_* \gg \vp_c$
which implies $\lambda^2 \gg \xi \sim 10^{-5}$. For smaller couplings
the full potential should be taken into account; this limit is
discussed below, at the end of the subsection. The power series expansion used
to derive \eref{expansion} is valid if $\lambda^2 (1+\alpha)$ is
small, so that the moduli corrections are subdominant. In general this
is also the condition for $n_s$ to be compatible with WMAP (an
exception to this is discussed in subsection~\ref{ssec:Bcrit}).
Using \eref{nsmod}, and \eref{alpha2},  we see that if our model is to
have any hope of compatibility with the CMB, it must satisfy
\be
10^{-5} \ll \lambda^2 \lesssim
\frac{1}{\alpha |\ln \lambda^2 g^2/Q^2|}
\sim \frac{V_0}{m_{3/2}^2|\ln \lambda^2  g^2/Q^2|} \, .
\label{lcons}
\ee
We note that this is also consistent with $\alpha \xi$ being small, as
is required for inflation to end in the right vacuum~\eref{alpha3}.
In contrast to $\lambda$, the gauge coupling $g$ is not strongly constrained.
This is not true for string motivated models for which the
Bogomolnyi bound applies, since $g \sim \lambda$. Substituting this,
and the definition of $V_0$, in to the above constraint, we obtain
\be
m_{3/2} \lesssim \xi \, .
\ee
Since $\xi \sim 10^{-5}$, this places a strong restriction on
$m_{3/2}$, which is equal to the moduli stabilisation scale for the
model we are studying. This implies that the inflation and moduli
scales must be dangerously close, unless the gauge coupling $g$ is
very small. This raises questions about the viability of such string
motivated inflation models. Even if $g$ is small, the above bound will
impose a strong restriction on the moduli sector. For the example
described by \eref{Wpar}, we find that $N > 206$ is needed to
satisfy the bound. However it should be noted that our moduli
stabilisation sector can hardly be described as generic, so perhaps
the problem can be evaded without changing the inflation sector significantly.

The usual treatment of hybrid inflation in a stringy context is to
assume the moduli are fixed at some high scale, and then to neglect
their presence during inflation. We see from \eref{lcons}
that this is only a good
approximation for a limited range of the superpotential coupling
$\lambda$.  Paradoxically enough, the larger the modulus scale and
thus the better fixed the moduli are, the less their
presence can be ignored, and the stronger the bound on $\lambda$
is.  For larger couplings the moduli corrections dominate, generically
resulting in a spectral index at odds with the WMAP result.

The above analysis using the asymptotic approximation~\eref{Nint}
will not be valid if inflation takes place close to the critical point
$\vp_* \approx \vp_c$. We will now study this limiting case, which
will apply for small coupling
$\lambda^2 \lesssim 10^{-5}$.  The modulus corrections are small if
$\lambda^2(1+\alpha) \ll 1$, which is automatic in this regime (it
follows from $\alpha \xi \lesssim 0.1$, see \eref{alpha3}, together with $\xi
\lesssim 10^{-5}$).  Hence, the usual SUSY results are recovered. In
particular, in the limit $\vp_* \to \vp_c$ we find
\bea
\xi &\approx & 10^{-5} \left( \frac{\lambda^2}{10^{-5}} \right)^{1/3}
\label{smallfield1}\\
n_s  &\approx& 1+ 2 \eta
\approx 1 +\frac{(\vp_*/\vp_c-1)}{8\pi^2} \ln(\vp_*/\vp_c-1)
\approx 1 \, .
\label{smallfield2}
\eea
The FI-term $\xi$ scales with $\lambda^{2/3}$ and can be made
arbitrary small for arbitrary small coupling.  However, in the small
coupling regime the perturbation spectrum is indistinguishable from a
scale invariant spectrum, which is disfavoured by the latest WMAP results.

\subsection{Large deviations from SUSY hybrid inflation}

We will now consider regions of parameter space in which the influence of the
moduli sector is significant. Typically this gives unacceptably large
corrections to $n_s$, allowing the corresponding parameters to be
ruled out.

As was shown in the previous section, if $\vp_*$ is close $\vp_c$,
then moduli corrections must always be small. Therefore we can take
the $\vp_* \gg \vp_c$ limit throughout this subsection, and use the
asymptotic expression~\eref{Nint} to find the behaviour of the inflaton.
Note that it is still possible to get $N_* \sim 60$ $e$-folds of
inflation no matter the value of $\mathcal{B}$. If $\mathcal{B} \geq
1$, the potential has no maximum and $N$ can be made arbitrary large
by taking the initial value of $\vp$ arbitrary
larger. Alternatively if $\mathcal{B}< 1$ tuning $\vp$ to start
sufficiently close to the maximum increases $N$.

The equation~\eref{Nint} can be analysed numerically, as we will
show in the next subsection. Although we cannot find analytic
solutions to it, we can find approximate analytic expressions in the
limiting cases of $\mathcal{B} \approx 1$ and $\mathcal{B} \ll 1$,
which are covered in subsections \ref{ssec:Bcrit} and \ref{ssec:max}
respectively. These results help us understand the
parameter dependence of the corrections to $n_s$. It should be noted that
the value of $n_s$ in these regimes is very sensitive to the effects of
the moduli sector. The renormalisation of the theory also needs to be
dealt with. For the model to be compatible with WMAP in these cases,
significant fine-tuning of the physics above the inflationary scale is
required.

\subsubsection{Numerical analysis}

Introducing the notation
\be
u
=\frac{g^2 N }{2\pi^2\vp_x^2 \mathcal{B}}
= \frac{\lambda^2(1+\alpha)}{6\pi^2} N
\, , \qquad
Y = \frac{\vp^2}{e \vp_x^2} \, ,
\ee
with $\vp_x$ and $\mathcal{B}$ given in~\eref{Bphix}, equation~\eref{Nint}
above can be brought into the form
\be
\frac{dY}{du}  = \frac{\mathcal{B}}{e} +  Y \ln Y \, .
\label{ueq}
\ee
The first term on the right is the usual $D$-term inflation potential,
whereas the $Y \ln Y$ term encodes the contribution from the moduli
sector. If this sector was absent the solution of the above equation
would reduce to $\mathcal{B} u = e Y$. In the $\vp \gg \vp_c$
asymptotic limit we are considering, $\eta \gg \epsilon$, and the
expression for the spectral index~\eref{nsconstr} reduces to
\be
n_s \approx 1+2\eta \approx 1 -
\frac{u}{N_* e Y}\left(\mathcal{B}- e Y[2+ \ln Y]\right) \, .
\label{nsu}
\ee
The COBE constraint~\eref{COBE} on the power spectrum becomes
\be
\xi =\frac{5 \sqrt{3} \delta_H }{2\sqrt{N_*}}
(\mathcal{B}+ e Y \ln Y)\left(\frac{u}{e \mathcal{B} Y}\right)^{1/2} \, .
\ee
If we take the limit $Y,u \to 0$ with $u \mathcal{B}/Y \to e$ in the
above expressions, they reduce to the standard SUSY hybrid inflation
results.

\begin{figure}
\centerline{ \includegraphics[width=3.7in]{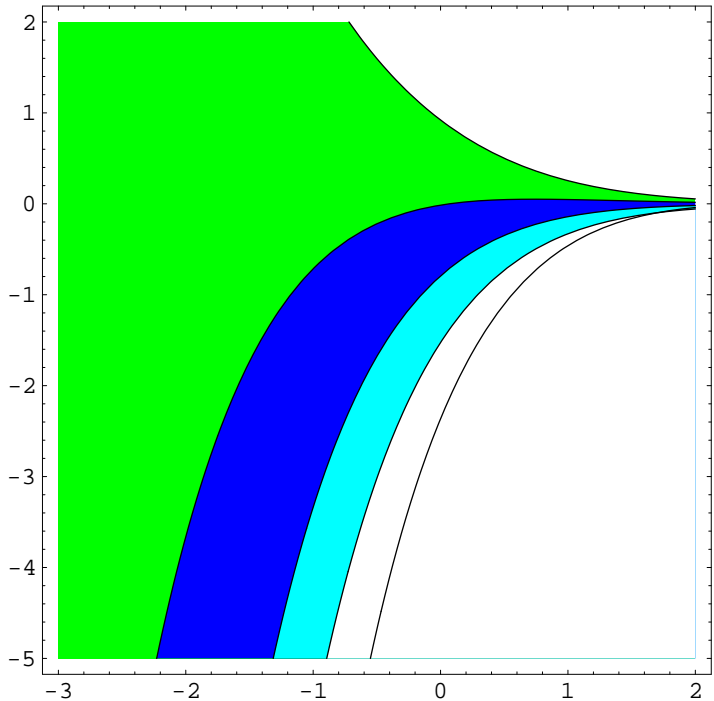} }
\caption{Contour plot of $n_s$ for $\ln \mathcal{B}$ vs.\ $\ln u$.
The upper contour is $n_s=1$, and the lower one is $n_s=0.9$. The
1-$\sigma$ WMAP3 results correspond to the dark (blue) region in between.
Parameters to the right of the black line below the contours satisfy
the cosmic string bound on $\xi$.}
\label{fig:ns}
\end{figure}

Having determined $Y$ numerically with \eref{ueq}, we see that the
spectral index $n_s$, and also $\xi$, depend only on the parameter
combinations $u$ and $\mathcal{B}$. The variation of $n_s$ in terms of
these parameters is plotted in \fref{fig:ns}. We see that $n_s$ is
compatible with WMAP only if $u$ is small or if $\mathcal{B}$ is close
to its critical value of one. The limit of small $u$, or equivalently
small $\lambda^2 (1+\alpha)$, corresponds to the corrections from the
moduli sector being subdominant. This region of parameter space was
covered analytically in subsection~\ref{ssec:SUSY}. For other
parameter choices the moduli corrections are significant. As we see
from \fref{fig:ns}, this usually leads to an unacceptable value of
$n_s$. An exception to this is when $B\approx 1$, where the different
corrections to $n_s$ cancel. This case is analysed in
subsection~\ref{ssec:Bcrit}.  For large $\alpha$, which is the most
realistic situation, $\mathcal{B}$ is small, and only the lower part
of figure~\ref{fig:ns} is relevant. An acceptable spectral index
typically requires $u$ small, which occurs when $\lambda$ is
small. This case is discussed in subsection~\ref{ssec:max}.

We note that since $u$ is proportional
to $\lambda^2$ and $\mathcal{B}$ to $g^2$ the $\lambda =\sqrt{2} g$ Bogomolnyi
limit is contained in \fref{fig:ns} as a $45^\circ$ straight line. The
position of the line depends on $\alpha$. For large $\alpha$, suitable
$n_s$ will only be obtained if $\lambda$ is very small or if $\mathcal{B}$ is
fine-tuned to be close to 1. This can place strong restrictions on the
string theory motivated inflation models for which this limit applies.

The value of $\xi$ required to satisfy the COBE bound can also be
found in terms of $u$ and $\mathcal{B}$. For the parameter choices
which give an acceptable value of $n_s$, we find $\xi$ in the range
$10^{-5}$--$10^{-6}$, so the moduli corrections do not change it
significantly. In particular we find that we cannot simultaneously
pull $\xi$ down to a value which will give a sufficiently low value of
the string tension $\mu$, and also keep $n_s$ in the acceptable
range. This can been seen from \fref{fig:ns}, where the cosmic string bound
on $\xi$ is represented by the black line. Hence some other mechanism
will be required to reduce the cosmic string contribution to the CMB.

\subsubsection{Approximation for $\mathcal{B} \approx 1$}
\label{ssec:Bcrit}

There is one regime where the moduli corrections can dominate, and yet
the spectral index remains in agreement with the data.  This is when
$\mathcal{B} \approx 1$, i.e.\ close to the critical value where the maximum
in the potential disappears.  To see how this comes about it would be
useful to have an analytic expression which covers this region, but
unfortunately we do not have an analytic solution for the
integral~\eref{Nint}. However, it
can be evaluated if we take
\be
-\frac{\vp^2}{\vp_x^2} \ln \frac{\vp^2}{e\vp_x^2}
\approx \left(2-\frac{\vp^2}{\vp_x^2}\right)
\frac{\vp^2}{\vp_x^2}
\ee
which is a reasonable approximation in the region $\vp^2/\vp_x^2 \lesssim e$.
Inflation in the small $\lambda^2 (1+\alpha)$ regime has already been
covered in subsection~\ref{ssec:SUSY}. Here we will take the
$\lambda^2 (1+\alpha) \gg 1$ limit, which will only realistically occur when
$\alpha \gg 1$ (since $\lambda \lesssim 1$).

In addition to the limit $\lambda^2 \alpha \gg 1$, we also take
$|\mathcal{B} -1| \ll 1$, since our numerical analysis tells us that
these are the only values of $\mathcal{B}$ which
will give a sensible value for $n_s$. The approximate inflaton solution
for these limits is
\be
\frac{\vp^2}{\vp_x^2} \approx
1 - \frac{6 \pi^2 N}{\lambda^2 \alpha}
+\left[\frac{\lambda^2 \alpha}{18\pi^2 N} +  \frac{1}{3}\right]
(\mathcal{B}-1) + \cdots \, .
\ee
Substituting this into the expression~\eref{nsu} for the spectral
index (and using the above approximation for $\vp^2 \ln \vp/\vp_x$) gives
\bea
\fl n_s \approx 1 - \frac{2}{N_*}\left[1+\frac{3 \pi^2 N_*}{\lambda^2 \alpha}
\right]
+ \frac{\lambda^4 \alpha^2(\mathcal{B}-1)}{54 \pi^4
  N_*^3}\left[ 1 +\frac{3\pi^2 N_*}{\lambda^2 \alpha}\right]
\nonumber \\ 
{}+ \Or\left(\frac{1}{\lambda^4\alpha^2},\mathcal{B}-1,
\lambda^8\alpha^4[\mathcal{B}-1]^2\right) \, .
\eea
So if we keep $\lambda^4 \alpha^2 |\mathcal{B}-1| \lesssim 1$, the
spectral index $n_s$ will remain in the acceptable range. This is in
qualitative agreement with our numerical analysis. However it should
be noted that $\mathcal{B} \sim g^2/[\alpha Q^2]$, so if
$\alpha$ is too large, then (since $g \lesssim 1$) it will not be
possible to have $\mathcal{B} \approx 1$. The solution in this
regime is extremely  sensitive to the moduli effects, and so its
viability requires fine-tuning of the high energy physics at
the moduli stabilisation scale. On a more positive note, we see from
\eref{Bmax} that if $\alpha \lesssim 18$, then it is still possible to
obtain agreement with the WMAP data even when $g, \lambda \sim
1$. Hence it may be possible for the Bogomolnyi limit to apply without
a very small value of $g$ being required.

\subsubsection{Low $\vp$ maximum ($\mathcal{B} \ll 1$)}
\label{ssec:max}

There is one other physically interesting region of parameter space
for which we can determine the approximate analytic behaviour of the inflaton.
We will consider the limiting case  where $V$ has a
maximum which occurs at a value of $\vp$ which is far lower than
$\vp_x$. This case corresponds to the  $\mathcal{B} \ll 1$ limit.
Since we naturally expect $\alpha$ to be large, and
$\mathcal{B} \sim 1/\alpha$, the above limit will frequently be satisfied.

When the potential~\eref{Vinf} has a maximum (i.e.\ if $\mathcal{B} \leq 1$),
we can rewrite the equation~\eref{Nint} as
\be
\frac{g^2}{2\pi^2} \frac{dN}{d\vp^2} =
\left[ 1 -\frac{\vp^2}{\vp_{\rm max}^2} -
\frac{\vp^2}{\vp_{\rm max}^2}
\frac{\ln (\vp^2/\vp_{\rm max}^2)}
{\ln (e\vp_x^2/\vp_{\rm max}^2)}
\right]^{-1} \, .
\ee
The magnitude of the third term in the bracket is bounded by $[e \ln
(e \vp_x^2/\vp_{\rm max}^2)]^{-1}$, so if
$\vp_x \gg \vp_{\rm  max}$, as is the case for small $\mathcal{B}$, we
can neglect it to leading order.
We then obtain the approximate solution
\be
\vp^2 \approx \vp_{\rm max}^2
+ \left(\vp_c^2 - \vp_{\rm max}^2\right)
\exp\left( -\frac{g^2 N}{2\pi^2 \vp_{\rm max}^2}\right)
\ee
which is valid throughout inflation. The usual SUSY result is obtained
in the limit $g^2 N/(2\pi^2), \vp_c^2 \ll \vp_{\rm max}^2$.

Taking $\vp \gg \vp_c$, and assuming $\epsilon \ll \eta$, we find
\be
n_s \approx 1 + 2\eta \approx 1 -\frac{g^2}{2\pi^2 \vp_{\rm  max}^2}
\left\{\left[1- \exp\left(
-\frac{g^2 N_*}{2\pi^2 \vp_{\rm max}^2}\right)\right]^{-1} + 1\right\}
\, .
\ee
We see that reducing $\vp_{\rm max}/g$ pushes
$n_s$ down, away from the usual SUSY result. If $\vp_*$ is too close
to the maximum, $n_s$ will be too low. This allows us to bound the
parameters of the moduli sector. For sensible values of $n_s$, we need
$g^2 N_*/[2\pi^2 \vp_{\rm max}^2]$ to be small. If
$\mathcal{B}$ is small, we can use the approximate
expression~\eref{aphimax} for $\vp_{\rm max}$. For
$\alpha \gg 1$, this implies the bound
\be
\lambda^2 \lesssim \frac{1}{\alpha \ln(\alpha Q^2/g^2)}
\, ,
\ee
which is similar to the bound derived in
subsection~\ref{ssec:SUSY}. Again, these
results are applicable in the Bogomolnyi limit.

We also find that the value of $\xi$ implied by the COBE constraint
decreases when $\vp_{\rm max}/g$ is reduced. This suggests that if
the moduli corrections are large enough, the cosmic string bound on
$\xi$ could be satisfied. However if we use the above bound on
$\vp_{\rm max}/g$ from the spectral index constraint, we find that
we must have $\xi \gtrsim 10^{-6}$, which does not satisfy the cosmic
string bound.

\section{Cosmic strings}
\label{sec:str}

In our model of $D$-term inflation coupled to moduli, cosmic strings
form during the breaking of the Abelian gauge symmetry $U(1)_{1}$ by
the waterfall field $\phi^+$ at the end of inflation.  The cosmic
strings in our model are not BPS as in usual $D$-term inflation.  This
is due to the non-vanishing of the gravitino mass and the coupling of
the string fields to the moduli sector. Non-observation of cosmic
strings constrains the string tension.  To see what this implies for
the parameters in our model, we will study the string solution and the
fermionic zero modes.

The  cosmic string solution has the form
\be
\phi^+ = [(2T_{\rm R} - |\chi|^2)\xi]^{1/2} e^{in\theta} f(r) \, ,
\qquad \phi^{-} = \phi =0 \, , \qquad
A_{\theta} = n a(r)
\ee
and the moduli sector fields depend only on $r$. The functions
$f(r)$, $a(r)$, are similar to the standard ones appearing in the
Abelian string model~\cite{ShelVil}. The Higgs field tends to its
vacuum value at infinity, and regularity implies it is zero at the
string core. Without loss of generality we can take the
winding number positive $n>0$.  If the variation of the moduli fields
inside the string is small, the string solution will be similar to a
BPS $D$-string, with a similar string tension, which (for $n=1$) is
\be
\mu \approx 2\pi \xi \, .
\ee
The above result will receive corrections from the one loop
contribution to the potential. We also expect that the variation of
the moduli fields inside the string will reduce $\mu$ (recall that
$\xi = \xi(T,\chi)$, so $\xi$ could be reduced inside the string
core). Throughout this paper we have assumed that the moduli are
stabilised by a very steep potential, which is not significantly
changed by any inflation scale contributions. The cosmic strings have
the same energy scale as inflation so, for the parameters we have
considered, we expect the above result to hold.  It should also be
noted that an important ingredient in the determination of the tension
is the width of the Higgs profile, which is set by the inverse Higgs
mass.  Although supersymmetry is broken in the vacuum, the Higgs mass
does not receive a soft mass contribution.  This is due to our choice
of K\"ahler potential, with the Higgs field inside the logarithm
\eref{noscale}.

The string tension is constrained by the data: CMB and pulsar timing
experiments give a comparable bound $G\mu \lesssim
10^{-7}$~\cite{pogosian,Lommen,ShelVil}.  This translates in a tight
constraint on the effective FI parameter $\xi$.  As discussed in
section 5, $\xi$ determines the size of the density perturbations
produced and is thus fixed by the CMB data. In most of parameter space
$\xi \approx 10^{-5}$ needed, in conflict with the cosmic string
bound.  The only exception is for very small superpotential couplings
$\lambda^2 \lesssim 10^{-9}$, but at the cost of having a slightly
disfavoured spectral index $n_{s} = 1$
(\ref{smallfield1},~\ref{smallfield2}).  It is interesting to note
that the string contribution can be small even when the gauge coupling
constant $g$ is large, this is because we are using a shift symmetry
for the inflaton. This is in sharp contrast to standard $D$-term
inflation which uses a minimal K\"ahler potential for the inflaton.
Another, more attractive, way to evade the cosmic string bound has
been proposed in~\cite{UAD}. If the model is augmented by a second set
of Higgs fields, which are put in a global $SU(2)$ multiplet together
with the original Higgs fields, the strings formed at the end of
inflation are semi-local~\cite{semilocal}. In the language of brane
inflation this requires an additional D7 brane~\cite{D3D7}. Semi-local
strings are not topologically stable, and they ultimately decay,
thereby easing the cosmic string bound.

Even if the string bound is somehow evaded, there are also constraints
arising from vorton bounds~\cite{BCDT,CD}, although these are very model
dependent~\cite{JP}. If a fermion couples to a string Higgs field, the
string can act as a potential well, and zero mode fermion bound states
can exist on it. Excitation of these will produce
currents~\cite{witten}. If the currents are stable and their
generation is not suppressed the vortons will be stable and will soon
dominate the energy density of the universe.  This potentially
constrains any model with fermion zero modes.

Since our model is supersymmetric, it will inevitably contain a large
number of fermion fields which provide candidates for zero mode bound
states~\cite{DDT}.  Given the potentially serious implications arising from
them, it is important to determine the number of zero modes.  The
number of zero mode solutions is given by an index theorem, and
depends only on the charges of the fermion under the string generator
(in our model the generator of the $U(1)_{1}$ symmetry) and the string
winding number~\cite{index1,index2}.   Applying the theorem to the
$\tilde{\phi}$-$\tilde{\phi}^{-}$ sector\footnote{Use (3.31) of
\cite{index2}, with effective charges $q_\phi =0$, $q_- = -n$.}
(which decouples from the other fermion fields), we find~\cite{RMstr}
\begin{equation}
N^-=2n
\label{zmm}
\end{equation}
negative chirality real zero modes (or equivalently $n$ complex
ones). There are $N^+=0$ positive chirality ones in this sector.

The other fermion fields couple to the gravitino, which complicates
things. Firstly, gravitinos are gauge fields, so it is necessary to
fix the gauge before analysing the system. In~\cite{index2} the
$\bar \sigma^\mu \psi_\mu =0$ gauge was used. This leaves three degrees of
freedom, which were denoted $\Psi$, $\Sigma$ and $\Pi$ in that
paper. Another complication arises from the fact that gravitinos have
different kinetic terms to the other fermions.  Nevertheless an
index theorem can be derived~\cite{index2}.  Applying it to our
model\footnote{Use section 6.2 of \cite{index2}, with effective
charges $q_\Psi = \mp 1$, $q_\Pi = q_\Sigma = \pm 1$, $q_+ = n$.}
gives $N^+ = 2(n+1)$, and $N^-=0$. However this result should be treated
with caution. The fermion norm obtained for the above gauge choice
does not appear to be positive definite, and so some of these apparent
zero mode solutions may be gauge. We conjecture that one complex zero
mode is due to a residual unbroken supersymmetry or Lorentz
transformation, and that the field $\Pi$ (see~\cite{index2}) should be
ignored. We then find
\begin{equation}
N^+=2n \, .
\label{zmp}
\end{equation}
Hence we find that there are equal numbers of positive and negative
chirality zero modes. Since the direction of the corresponding
massless states is determined by their chirality, we see there are
equal number of left and right movers on the string. This could lead to
a smaller net current and hence weaker bounds on the model.

It is interesting to contrast the above results with those for
$D$-term inflation without a moduli sector. The number of negative
chirality zero modes is unchanged (irrespective of whether the model
is SUSY or SUGRA). In a SUSY model the number of positive chirality
modes is $2n$~\cite{DDT}, just like \eref{zmp}. However when the gravitinos are
included (which will be massless when the moduli sector is absent),
they mix with one pair of zero modes and it is no longer a bound
state~\cite{RMstr}, so $N^+$ is reduced to $2(n-1)$~\cite{index2}. Hence for a
winding number $n=1$ string, all zero modes move in the same direction
and the current will be maximal (and the bounds stronger)~\cite{Dlett}. On the
other hand, for the model described in this paper, the introduction of
the moduli sector gives a non-zero gravitino mass. This helps
relocalise the zero mode, and the resulting currents are no longer chiral.

If vortons do arise in our model then the string tension will be
constrained since vortons are long lived. There are two constraints
one can apply~\cite{BCDT,CD}. Firstly, and more conservatively, the
universe should be radiation dominated at nucleosynthesis. Consequently,
this constrains the string tension such that $G\mu \lesssim 10^{-20}$.
Alternatively if the vortons are very long lived then we require
them not to overclose the universe, in which case $G\mu \lesssim 10^{-24}$.
These constraints are much stronger than the current bounds
on cosmic $D$-strings in brane inflation models, but are not as strong
as those cited in~\cite{Dlett} where only BPS strings were
considered. A possible way around this is for there to be a low reheat
temperature. In this case we could envisage the possibility of
$D$-loop dark matter. We note that semi-local strings are likely to evade
these bounds since they are probably not stable enough for vorton formation.

\section{Conclusions}
\label{sec:conc}

Hybrid inflation can be realised in supergravity as the low energy
description of string systems such as the D3/D7 brane
configurations. At tree level the inflation potential is flat, only
lifted by loop corrections leading to a slow rolling inflationary
phase. However it must not be forgotten that in the full underlying
theory, there will be other non-inflationary fields, which could
interfere with slow-roll. In particular, moduli fields are pervasive
in most string compactifications. If not stabilised, these lead to
flat directions at the perturbative level. This poses
a serious problem for any attempt to embed inflationary
cosmology within string theory. Even if the moduli are stabilised, the
stabilisation mechanism itself can damage the flatness of the
inflationary potential.

In this paper we have analysed a combined SUGRA model of $D$-term hybrid
inflation and stabilised moduli. The inclusion of moduli requires
substantial changes to most aspects of the inflation model. First of all,
it is difficult to use a constant Fayet-Iliopoulos term to give the
inflation vacuum energy when other sectors are present in the theory.
Instead, we have shown how an effective FI-term can be obtained from
the VEVs of stabilised moduli fields. The moduli sector we took has an
additional Abelian gauge symmetry. For suitable moduli charges, the
resulting $D$-term is positive definite. This second $D$-term allows
one to obtain an effective Fayet-Iliopoulos term, whose presence is
required to lift the AdS vacuum in the moduli sector to a Minkowski
one. By coupling the moduli fields to the inflationary $U(1)$, this
effective Fayet-Iliopoulos term can also be made to appear in the
inflationary $D$-term, making inflation possible.

The inclusion of the moduli sector also has implications for the
choice of K\"ahler potential. Using a minimal K\"ahler potential for
the inflationary sector results in a large mass for the inflaton
field, destroying the flatness of the potential. In other words, the
presence of the moduli sector causes the well-known $\eta$-problem to
appear in $D$-term inflation. The problem can be cured by introducing
a shift symmetry for the inflaton field in the K\"ahler potential. The
inflaton potential is then obtained at the one loop level, but
receives drastic modifications due to the coupling to moduli. The
requirement that inflation ends normally imposes further restrictions
on the full K\"ahler potential. Taking it to be of the no-scale form
\eref{noscale} gives a working model.  The variation of moduli fields
during inflation is another potential source of problems, although for
our $D$-term model they can easily be avoided.  

As was discussed in section~\ref{sec:inf}, the moduli corrections
produce inflationary potentials with a variety of forms. Many of these are not
compatible with observations. A parameter space analysis was
performed, imposing both the COBE normalisation and the WMAP
constraints on the spectral index. Three acceptable regions of
parameter space are summarised in Table~\ref{table}.
In contrast to $D$-term inflation without moduli, one can accommodate
a spectral index near one here. Lower values, closer to the favoured
WMAP3 value are also possible. We find that compatibility of the
spectral index with WMAP generally requires that the moduli
corrections to $D$-term inflation are very small. This can be achieved
by taking the inflation superpotential coupling $\lambda$ to be small
(columns A and B in the table). There is also a region of parameter
space which can give reasonable $n_s$ even if $\lambda$ is big (column
C of the table). It corresponds to the different effects of the moduli
sector cancelling, and (not surprisingly) requires fine-tuning of this
sector.

There are additional constraints on the ratio 
$\alpha \sim m_{3/2}^2/(g^2 \xi^2)$. Note that the moduli
stabilisation sector used in this paper has just one mass scale, which
is also the gravitino mass. Since the moduli fields are stabilised
before inflation, this implies that $m_{3/2}$ is generally
large. Combining this with the requirement that the moduli sector corrections
do not prevent inflation from ending, implies 
$1 \lesssim \alpha \lesssim \xi^{-1}$. Satisfying all of the above
constraints simultaneously is particularly difficult for the Bogomolnyi case 
$g \sim \lambda$. It requires a very small value of $g$ and a fine-tuned
moduli sector, neither of which are natural.

\begin{table}[t]
\begin{center}
\begin{tabular}[h]{|c||c|c|c|}
\hline & A & B & C \\ \hline 
\hline $\lambda$ \rule{0cm}{0.5cm} & $10^{-5} \lesssim
  \lambda^2 \lesssim {g^2 \xi^2 / m_{3/2}^2}$ & $\lambda^2 \lesssim 10^{-5}$ &
  $1 \ll \lambda^2 {g^2 \xi^2 / m_{3/2}^2} \lesssim |\mathcal{B}-1|^{-1/2}$ \\
\hline $\xi$ \rule{0cm}{0.5cm} & $10^{-5}$ &
 $10^{-5} (\lambda^2  /10^{-5})^{1/3}$ &  $10^{-6} - 10^{-5}$ \\ 
\hline $n_s$ \rule{0cm}{0.5cm} & $0.98 + \Or(\lambda^2 g^2
  \xi^2 / m_{3/2}^2 )$ & 1 & $0.96 + \Or( m_{3/2}^2/ \lambda^2 g^2
  \xi^2)$\\\hline
\end{tabular}
\end{center}
\caption{Parameter ranges for viable inflation. The gauge
coupling satisfies $g \lesssim 1$. The renormalisation scale-dependent
parameter $\mathcal{B}$ is defined in subsection~\ref{ssec:dV}, and
$1 \ll \alpha \ll\xi^{-1}$ has been assumed.
}
\label{table}
\end{table}

Finally, the structure of the cosmic strings produced at the end of
inflation is different. They are no longer
BPS objects, due to the SUSY-breaking produced by the moduli sector.
The string contribution to CMB is problematic as in the SUSY $D$-term
case~\cite{prd}, unless $\lambda$ is very small in which case
the influence from the moduli sector is negligible and we recover the
SUSY $D$-term regime~\cite{infl,mairi}. This conundrum can be resolved by
resorting to a doubling of the charged fields corresponding to the
presence of a second D7 brane, and the formation of metastable semi-local
strings~\cite{semilocal}.
We should stress that the cosmic string bound is not specific to our
model (all models of $D$-term inflation face this problem), and in
particular that it is almost independent of the moduli sector.

The number of fermionic zero modes on a winding number $n$ string
is also altered, being $n$ complex fermions of both chiralities
here compared to just $n-1$ chiral modes in the BPS
case~\cite{RMstr,index2}. These can potentially cause further
conflicts with cosmology. Although it should be noted that the
corresponding bounds are sensitive to the trapping possibilities
of these zero modes, as well as their survival on cosmic string
loops. The stability of the possible vorton configurations is also
a factor. This will be less significant for the semi-local cases.

In summary, we have seen that moduli fields cannot be treated
lightly in the context of inflationary cosmology. They lead to
observable differences in both the inflation and the topological
defect sectors of the theory. In a wide range of cases the differences are
serious enough for the models to be ruled out. There is
still a large region of parameter space in which inflation works,
despite the presence of moduli.
Nevertheless, notice that models compatible with the WMAP3 results lead to
the formation of cosmic strings with a large tension. To alleviate this
tension, one may introduce copies of the scalar fields leading to
semi-local strings. This is left for future work.

\ack We are grateful to D. H. Lyth, A. Achucarro, C. Burgess and
S. Vandoren for helpful comments and discussions.  ACD, SCD and CvdB
thank CEA Saclay for their hospitality. ACD is also grateful to the
Galileo Galilei Institute and INFN for hospitality and partial
support. For financial support, SCD and RJ thank the Netherlands
Organisation for Scientific Research (NWO), MP thanks FOM, and CvdB and
ACD thank PPARC for partial support.  PhB
acknowledges support from RTN European programme MRN-CT-2004-503369.

\end{document}